\documentclass[10pt]{article}
\pdfoutput 1
\usepackage{graphicx}
\usepackage{amsmath}
\usepackage{amssymb}
\usepackage[colorlinks=true]{hyperref}
\usepackage{pifont}% http://ctan.org/pkg/pifont

\DeclareFontFamily{OT1}{pzc}{}
\DeclareFontShape{OT1}{pzc}{m}{it}{<-> s * [1.10] pzcmi7t}{}
\DeclareMathAlphabet{\mathpzc}{OT1}{pzc}{m}{it}

\newcommand{\vphi}{\varphi}

\newcommand{\bs}[1]{\ensuremath{\boldsymbol{#1}}}

\newcommand{\Jphi}{\ensuremath{J_\varphi}}
\newcommand{\verphi}{\ensuremath{\bs{\hat{\varphi}}}}
\newcommand{\verx}{\ensuremath{\bs{\hat{x}}}}
\newcommand{\very}{\ensuremath{\bs{\hat{y}}}}

\title{The vector potential of a steady azimuthal current
  density.\\Once again.}  \author{Antonio O.\ Bouzas\thanks{email:
    abouzas@cinvestav.mx.
    ORCID: 0000-0001-5493-4958.
  }  \\\small
  Departamento de F\'{\i}sica Aplicada, CINVESTAV-IPN \\[-4pt]\small
  Carretera Antigua a Progreso Km.\ 6, Apdo.\ Postal 73
  ``Cordemex''\\[-4pt]\small M\'erida 97310, Yucat\'an, M\'exico}
\date{22 September 2025}

\begin{document}

\maketitle
\begin{abstract}
  We give an integral expression for the vector potential of a
  time-independent, steady azimuthal current density. Our derivation
  is substantially simpler and somewhat more general than others given
  in the literature. As an illustration, we recover the results for
  the vector potential of a circular current loop as an orthogonal
  expansion in spherical and cylindrical coordinates. Additionally, we
  obtain closed analytical expressions for the vector potential and
  the magnetic induction of a circular current loop in terms of
  Legendre functions of the second kind, that are simpler than the
  results in terms of complete elliptic integrals given in textbooks.
\end{abstract}

%\tableofcontents{}

\section{Introduction}
\label{sec:intro}

We consider a magnetostatic \cite{jack,zang} system defined by a known time-independent
current density $\bs{J}(\bs{r})$ which is in a steady state,
\begin{equation}
  \label{eq:intro.1}
  \bs{\nabla}\cdot\bs{J}(\bs{r}) =0.
\end{equation}
We further assume, for simplicity, that $\bs{J}(\bs{r})$ is spatially
localized, that is, there exists $R>0$ such that $\bs{J}(\bs{r})=0$ if
$|\bs{r}|>R$. Under these conditions, the vector potential
$\bs{A}(\bs{r})$ in Coulomb gauge,
\begin{equation}
  \label{eq:intro.2}
  \bs{\nabla}\cdot\bs{A}(\bs{r}) =0,  
\end{equation}
satisfies the vector Poisson equation with boundary conditions given by,
\begin{equation}
  \label{eq:intro.3}
  \bs{\nabla}^2 \bs{A}(\bs{r}) = -\mu_0 \bs{J}(\bs{r}),
  \qquad
  \lim_{|\bs{r}|\to\infty} \bs{A}(\bs{r}) =0~.
\end{equation}
From these equations it is clear that if $\bs{J}$ points in a fixed
direction, $\bs{J}=J \bs{\hat{J}}$, then $\bs{A}$ must be parallel to
$\bs{J}$, $\bs{A}=A_J \bs{\hat{J}}$. Indeed, for any unit vector
$\bs{\hat{k}}$ satisfying $\bs{\hat{k}}\cdot\bs{J}=0$, we obtain from
(\ref{eq:intro.3}) that $\bs{\hat{k}}\cdot\bs{A}$ satisfies the
Laplace equation with a homogenous boundary condition at infinity,
and must therefore vanish: $\bs{\hat{k}}\cdot\bs{A}=0$.

If the current density points in the azimuthal direction,
\begin{equation}
  \label{eq:intro.4}
  \bs{J}(\bs{r})=J_\vphi(\bs{r}) \verphi,
  \qquad
  \verphi=-\sin(\vphi) \verx + \cos(\vphi) \very  
\end{equation}
with $\vphi$ the azimuthal coordinate in spherical or cylindrical
coordinates,  the 
steady-state condition (\ref{eq:intro.1}) yields
\begin{equation}
  \label{eq:intro.5}
\frac{\partial J_\vphi}{\partial\vphi}(\bs{r})=0.
\end{equation}
In this case, because the unit vector $\verphi$ is not constant, it is
not obvious that $\bs{A}$ is parallel to $\bs{J}$. Here, we show that
for a stationary current density of the form (\ref{eq:intro.4}),
satisfying (\ref{eq:intro.5}), we have
\begin{equation}
  \label{eq:intro.6}
  \bs{A}(\bs{r}) = A_\vphi(\bs{r}) \verphi,
  \qquad
    A_\vphi(\bs{r}) = \frac{\mu_0}{4\pi} \int d^3r' 
  \frac{\cos(\vphi'-\vphi)}{|\bs{r}-\bs{r}'|} \Jphi(\bs{r}').  
\end{equation}
Several comments on this equation are in order. First, on the $z$ axis
the denominator becomes independent of $\vphi'$ and, therefore, the
integral vanishes: $A_\vphi(z\bs{\hat{z}}) = 0$. Second, the
denominator depends on $\vphi'$ only through $\vphi'-\vphi$, like the
numerator, which means that we could set $\vphi$ to any fixed value in
the integrand without changing the integral. Thus, $\bs{A}(\bs{r})$
does not depend on $\vphi$. The equality (\ref{eq:intro.6}) with
$\vphi=0$ is proven in spherical coordinates in \cite{vas12}, and in
cylindrical coordinates in section 10.5.3 of \cite{zang}. The proof
given below is substantially simpler than that in \cite{vas12}, and
more general than the proofs in both \cite{zang,vas12}, as it is valid
in spherical and cylindrical coordinates and, in fact, in any
orthogonal system of curvilinear coordinates containing the azimutal
angle $\vphi$. Below, we apply (\ref{eq:intro.6}) to recover the known
results for the vector potential of a circular current loop as an
expansion in spherical harmonics, and as an integral transform of
Bessel functions in cylindrical coordinates.  We use the latter form,
in turn, to find a new, simpler exact solution in closed analytical
form for the vector potential and the magnetic induction of the
circular current loop in terms of Legendre functions of the second
kind, that are simpler than the results in terms of complete elliptic
integrals given in textbooks.

%Because the arguments advanced below are elementary, they may have
%been found many times in the past. The purpose of this note is
%therefore not so much to claim originality as to provide reading
%material for students of electrodynamics.
This note is intended as reading material for graduate or advanced
undergraduate students, to complement the treatment of magnetostatics
in textbooks such as \cite{jack,zang}.  The paper is organized as
follows. In the next section, we provide an elementary proof of
(\ref{eq:intro.6}) that is valid for any coordinate system containing
the azimuthal angle $\vphi$. In section \ref{sec:examp} we discuss the
well-known example of the vector potential of a circular current loop
as an orthogonal expansion in spherical (section \ref{sec:examp.sph})
and cylindrical (section \ref{sec:examp.cyl}) coordinates.  We also
discuss closed analytical forms for the vector potential and magnetic
induction in section \ref{sec:examp.analytic}, where we derive a
simpler expression than those found in textbooks. Finally, in section
\ref{sec:finrem}, we provide a brief summary

\section{Vector potential}
\label{sec:vec.pot}

The solution to the equations (\ref{eq:intro.3}) with the condition
(\ref{eq:intro.2}) is given by \cite{jack,zang},
\begin{equation}
  \label{eq:vecpot.0}
  \bs{A}(\bs{r})= \frac{\mu_0}{4\pi} \int d^3r' \frac{\bs{J}(\bs{r}')}{|\bs{r}-\bs{r}'|},
\end{equation}
From (\ref{eq:intro.4}), (\ref{eq:intro.5}) we then have,
\begin{equation}
  \label{eq:vecpot.1}
  \bs{A}(\bs{r})= \frac{\mu_0}{4\pi} \int dV_2' \Jphi(\bs{r}')
  \int_{-\pi}^{\pi} d\varphi' \frac{\verphi'}{|\bs{r}-\bs{r}'|},
\end{equation}
with
\begin{equation}
  \label{eq:vecpot.2}
  \int dV_2' =\left\{
  \begin{array}{ll}
\displaystyle   \int_0^\infty dr' r'^2\int_0^\pi d\theta' \sin(\theta') & \text{in spherical coordinates}\\[10pt]
\displaystyle    \int_0^\infty d\rho' \rho'\int_{-\infty}^\infty dz'     & \text{in cylindrical coordinates}
  \end{array}
\right..
\end{equation}
It will be useful below to define the notation
\begin{subequations}
\label{eq:vecpot.3}
\begin{equation}
  \label{eq:vecpot.3a}
  R(\alpha) = \left(r^2+r'^2-2 rr' \cos(\theta)\cos(\theta')-2 r r' \sin(\theta)\sin(\theta') \cos(\alpha) \right)^\frac{1}{2},
\end{equation}
in spherical coordinates, and 
\begin{equation}
  \label{eq:vecpot.3b}
  R(\alpha) = \left(\rho^2+z^2+\rho'^2+z'^2-2 z z'-2 \rho\rho' \cos(\alpha) \right)^\frac{1}{2},
\end{equation} 
\end{subequations}
in cylindrical ones. The function $R(\alpha)$ in (\ref{eq:vecpot.3}) has
three properties that will be needed below: (i) it is a
$2\pi$-periodic function of $\alpha$, (ii) it is an even function of
$-\pi<\alpha<\pi$, and (iii) in both spherical and cylindrical
coordinates it is true that
\begin{equation}
  \label{eq:vecpot.4}
  |\bs{r}-\bs{r}'| = R(\vphi'-\vphi).
\end{equation}
We can then rewrite (\ref{eq:vecpot.1}) as,
\begin{align}
  \label{eq:vecpot.5}
  \bs{A}(\bs{r}) &= \frac{\mu_0}{4\pi} \int dV_2' \Jphi(\bs{r}')
  \int_{-\pi}^{\pi} d\varphi'
  \frac{-\sin(\vphi')\verx+\cos(\vphi')\very}{R(\vphi'-\vphi)}.\\
\intertext{Let $\vphi''=\vphi'-\vphi$, then}
  &= \frac{\mu_0}{4\pi} \int dV_2' \Jphi(\bs{r}')
  \int_{-\pi+\vphi}^{\pi+\vphi} d\varphi''
    \frac{-\sin(\vphi''+\vphi)\verx+\cos(\vphi''+\vphi)\very}{R(\vphi'')}.
    \label{eq:vecpot.5a}
\end{align}
We now use the trigonometric relation
\begin{equation}
  \label{eq:vecpot.6}
  -\sin(\vphi''+\vphi)\verx+\cos(\vphi''+\vphi)\very =  -\sin(\vphi'')
  \bs{\hat{n}} + \cos(\vphi'') \verphi,
\end{equation}
with $\verphi$ as in (\ref{eq:intro.4}) and
\begin{equation}
  \label{eq:vecpot.7}
  \bs{\hat{n}} = \left\{
  \begin{array}{ll}
\displaystyle   \sin(\theta) \bs{\hat{r}} + \cos(\theta) \bs{\hat{\theta}} & \text{in spherical coordinates}\\[10pt]
\displaystyle   \bs{\hat{\rho}}                                            & \text{in cylindrical coordinates}
  \end{array}
\right.,
\end{equation}
to rewrite (\ref{eq:vecpot.5a}) as
\begin{align}
  \label{eq:vecpot.8}
  \bs{A}(\bs{r}) &= \frac{\mu_0}{4\pi} \int dV_2' \Jphi(\bs{r}')
  \int_{-\pi+\vphi}^{\pi+\vphi} d\varphi''
  \frac{-\sin(\vphi'') \bs{\hat{n}} + \cos(\vphi'') \verphi}{R(\vphi'')},
  \intertext{Because the integrand is $2\pi$-periodic in $\vphi''$, we have,}
  &= \frac{\mu_0}{4\pi} \int dV_2' \Jphi(\bs{r}')
  \int_{-\pi}^{\pi} d\varphi''
    \frac{-\sin(\vphi'') \bs{\hat{n}} + \cos(\vphi'') \verphi}{R(\vphi'')},
    \intertext{and because the term proportional to $\bs{\hat{n}}$ in the integrand is and odd function of $\vphi''$, its integral vanishes and we have,} 
&= \frac{\mu_0}{4\pi} \verphi \int dV_2' \Jphi(\bs{r}')
  \int_{-\pi}^{\pi} d\varphi''
    \frac{\cos(\vphi'')}{R(\vphi'')}.
\end{align}
We can now make the change of variable $\vphi''=\vphi'-\vphi$ to obtain,
\begin{align}
  \label{eq:vecpot.9}
  \bs{A}(\bs{r}) &= \frac{\mu_0}{4\pi} \verphi\int dV_2' \Jphi(\bs{r}')
  \int_{-\pi+\vphi}^{\pi+\vphi} d\varphi' \frac{\cos(\vphi'-\vphi)}{R(\vphi'-\vphi)}.
\intertext{By invoking the $2\pi$-periodicity of the integrand as a function of $\vphi'$, we get}
  \label{eq:vecpot.a}
  \bs{A}(\bs{r}) &= \frac{\mu_0}{4\pi} \verphi\int dV_2' \Jphi(\bs{r}')
  \int_{-\pi}^{\pi} d\varphi' \frac{\cos(\vphi'-\vphi)}{R(\vphi'-\vphi)},
\end{align}                   
and using (\ref{eq:vecpot.2}), (\ref{eq:vecpot.4}) we arrive at (\ref{eq:intro.6}).
%\begin{equation}
%  \bs{A}(\bs{r}) = \frac{\mu_0}{4\pi} \verphi\int d^3r' \Jphi(\bs{r}')
%  \frac{\cos(\vphi'-\vphi)}{|\bs{r}-\bs{r}'|}.  
%\end{equation}

\section{The circular current loop}
\label{sec:examp}

As a simple application of (\ref{eq:intro.6}), we consider the vector
potential of a planar circular current loop of radius $a$ and current
$I$ in spherical coordinates, as discussed in section 5.5 of
\cite{jack}, as well as in cylindrical ones, as discussed in Example
10.6 of \cite{zang}.

\subsection{Spherical coordinates}
\label{sec:examp.sph}

In spherical coordinates, the current density of a circular loop lying
at the origin on the $xy$ plane is given by
(\ref{eq:intro.4}) with
\begin{equation}
  \label{eq:examp.sph.1}
  J_\vphi = \frac{I}{a} \sin\theta' \delta(\cos\theta') \delta(r'-a).
\end{equation}
Substituting this value in (\ref{eq:intro.6}) we get,
\begin{equation}
  \label{eq:examp.sph.2}
  A_\vphi = \frac{\mu_0}{4\pi}aI\int_{-\pi}^{\pi} d\vphi' \left[
    \frac{\cos(\vphi'-\vphi)}{|\bs{r}-\bs{r'}|}
    \right]_{\hspace*{-6pt}
      \begin{array}{l}
        \scriptstyle r'=a\\[-4pt]
        \scriptstyle \theta'=\pi/2
      \end{array}
}.
\end{equation}
This integral is computed in \cite{jack} both in closed form in terms
of elliptic functions in eq.\ (5.37), and as an expansion in Legendre
associated functions in eq.\ (5.46). We consider the former in section
\ref{sec:examp.analytic} below, and in this section we briefly comment
on the latter. 

We employ the expansion of $1/|\bs{r}-\bs{r'}|$ in spherical
harmonics, as given in eq.\ (3.70) of \cite{jack} or (4.83) of \cite{zang},
%and expand the
%spherical harmonics themselves into an associated Legendre function of
%$\cos\theta$ and imaginary exponential of $\vphi$ as in eq.\ (3.53) of
%\cite{jack},
to obtain
%\begin{equation}
%  \label{eq:examp.sph.3}
%  A_\vphi = \frac{\mu_0}{4\pi}aI \sum_{\ell=0}^{\infty} \sum_{m=-\ell}^{\ell}
%  \frac{(\ell-m)!}{(\ell+m)!} \frac{r_{<}^{\ell}}{r_{>}^{\ell+1}}
%  P_\ell^m(0) P_\ell^m(\cos\theta)
%  \int_{-\pi}^{\pi} d\vphi' \cos(\vphi'-\vphi) e^{-i m (\vphi'-\vphi)}.
%\end{equation}
\begin{equation}
  \label{eq:examp.sph.3}
  \begin{aligned}
  A_\vphi &= \mu_0 a I \sum_{\ell=0}^{\infty} \sum_{m=-\ell}^{\ell}
  \frac{1}{2\ell+1} \frac{r_{<}^{\ell}}{r_{>}^{\ell+1}}
\int_{-\pi}^{\pi} d\vphi' \cos(\vphi'-\vphi) Y_{\ell,m}(\theta,\vphi)
Y_{\ell,m}^*(\pi/2,\vphi')\\
&= \mu_0 a I \sum_{\ell=0}^{\infty} \sum_{m=-\ell}^{\ell}
\frac{1}{2\ell+1} \frac{r_{<}^{\ell}}{r_{>}^{\ell+1}}
Y_{\ell,m}(\theta,0) Y_{\ell,m}(\pi/2,0)
\int_{-\pi}^{\pi} d\vphi' \cos(\vphi'-\vphi) e^{-i m (\vphi'-\vphi)}.
  \end{aligned}
\end{equation}
where $r_<=\min\{a,r\}$ and $r_>=\max\{a,r\}$ and in the second line
we used
$Y_{\ell,m}(\theta,\vphi)=Y_{\ell,m}(\theta,0)e^{im\vphi}$. The
integration over the azimuthal angle in (\ref{eq:examp.sph.3}) is
immediate
\begin{equation}
  \label{eq:examp.sph.4}
  \int_{-\pi}^{\pi} d\vphi' \cos(\vphi'-\vphi) e^{-i m (\vphi'-\vphi)}
  = \pi (\delta_{-1m} + \delta_{1m}).
\end{equation}
Substituting this in (\ref{eq:examp.sph.3}) and using
$Y_{\ell,-m}(\theta,0)=(-1)^m Y_{\ell,m}(\theta,0)$ yields
\begin{equation}
  \label{eq:examp.sph.6}
  A_\vphi = 2\pi \mu_0aI \sum_{\ell=1}^{\infty}
  \frac{1}{2\ell+1}\frac{r_{<}^{\ell}}{r_{>}^{\ell+1}}
  Y_{\ell,1}(\theta,0) Y_{\ell,1}(\pi/2,0)
\end{equation}
We can expand the spherical harmonics into an associated Legendre function of
$\cos\theta$ and imaginary exponential of $\vphi$ as in eq.\ (3.53) of
\cite{jack} or (C.18) of \cite{zang}, to obtain the equivalent form
%\begin{equation}
%  \label{eq:examp.sph.5}
%  A_\vphi = \frac{\mu_0aI}{4} \sum_{\ell=1}^{\infty}
%  \frac{r_{<}^{\ell}}{r_{>}^{\ell+1}} \left(
%    \frac{(\ell+1)!}{(\ell-1)!}  P_\ell^{-1}(0) P_\ell^{-1}(\cos\theta) +
%    \frac{(\ell-1)!}{(\ell+1)!}  P_\ell^{1}(0) P_\ell^{1}(\cos\theta)
%  \right).
%\end{equation}
%This expression can be simplified by using the relation
%$P_\ell^{-1}(x) = -P_\ell^{1}(x) (\ell-1)!/(\ell+1)!$, as given in
%eq.\ (3.51) of \cite{jack}, which yields
\begin{equation}
  \label{eq:examp.sph.7}
  A_\vphi = \frac{\mu_0aI}{2} \sum_{\ell=1}^{\infty}
  \frac{1}{(\ell+1)\ell} \frac{r_{<}^{\ell}}{r_{>}^{\ell+1}} 
      P_\ell^{1}(0) P_\ell^{1}(\cos\theta).
\end{equation}
This is our result for the vector potential of the circular current
loop. Since $P_\ell^{1}(0)$ vanishes for even $\ell$, the sum in
(\ref{eq:examp.sph.6}) runs only over odd values of $\ell$. Furthermore,
by using the expression for $P_\ell^{1}(0)$ given in eq.\ (5.45)
of \cite{jack}, we can show that (\ref{eq:examp.sph.6}) is
mathematically equivalent to the expression for the vector potential
given in eq.\ (5.46) of \cite{jack}. To obtain the magnetic field we
have to take the curl of (\ref{eq:intro.6}), (\ref{eq:examp.sph.6});
this has been explicitly done in \cite{jack} for the circular loop,
and in \cite{vas12} for the potential (\ref{eq:intro.6}). There is no
need to repeat those calculations here.

\subsection{Cylindrical coordinates}
\label{sec:examp.cyl}

In cylindrical coordinates, the current density of a circular loop lying
at the origin on the $xy$ plane is given by
(\ref{eq:intro.4}) with
\begin{equation}
  \label{eq:examp.cyl.1}
  J_\vphi = I \delta(z') \delta(\rho'-a).
\end{equation}
Substituting this value in (\ref{eq:intro.6}) we get,
\begin{equation}
  \label{eq:examp.cyl.2}
  A_\vphi = \frac{\mu_0}{4\pi}aI\int_{-\pi}^{\pi} d\vphi' \left[
    \frac{\cos(\vphi'-\vphi)}{|\bs{r}-\bs{r'}|}
    \right]_{\hspace*{-6pt}
      \begin{array}{l}
        \scriptstyle \rho'=a\\[-4pt]
        \scriptstyle z'=0
      \end{array}
}.
\end{equation}
We use now the expansion of $1/|\bs{r}-\bs{r'}|$ as a trigonometric
series of Laplace transforms given in Problem 3.16 of
\cite{jack}, which we repeat here for convenience,
\begin{equation}
  \label{eq:examp.cyl.3}
  \frac{1}{|\bs{r}-\bs{r'}|}
  =
  \sum_{m=-\infty}^\infty e^{-i m (\vphi'-\vphi)} \int_0^\infty dk
  J_m(k\rho') J_m(k\rho) e^{-k |z-z'|}.
\end{equation}
Substituting (\ref{eq:examp.cyl.3}) in (\ref{eq:examp.cyl.2}) and
using the angular integral (\ref{eq:examp.sph.4}) and the relation
$J_{-1}(x)=-J_{1}(x)$ yields, 
\begin{equation}
  \label{eq:examp.cyl.4}
  A_\vphi(\bs{r}) = \frac{\mu_0 I a}{2} \int_0^\infty dk J_1(ka)
  J_1(k\rho) e^{-k |z|}, 
\end{equation}
which expresses the vector potential (\ref{eq:intro.6}) as an integral
transform in cylindrical coordinates. Equation (\ref{eq:examp.cyl.4})
recovers the result in Example 10.6 in Section 10.5.4 of \cite{zang}.

\subsection{Analytically closed expressions: vector potential}
\label{sec:examp.analytic}

The integral in (\ref{eq:examp.cyl.4}) is a Laplace transform that can
be readily evaluated with the software system Mathematica \cite{wolf}
in analytically closed form in terms of complete elliptic integrals,
\begin{equation}
  \label{eq:examp.clo.1}
    A_\vphi(\bs{r}) =
    \frac{\mu_0 I}{2\pi}\frac{\sqrt{(a+\rho)^2+z^2}}{\rho}
    \left(\frac{a^2+\rho^2+z^2}{(a+\rho)^2+z^2} K(\kappa) - E(\kappa)
    \right),
    \quad
    \kappa = \sqrt{\frac{4 a\rho}{(a+\rho)^2+z^2}}.
\end{equation}
Here, $K$ and $E$ are the complete elliptic integrals of the first and
second kind, respectively, as defined in section 13.8 of
\cite{bates.ell} (see also \cite{wiki.ell} for quick reference). From
a practical point of view, we point out that in Mathematica one has,
\begin{equation}
  \label{eq:examp.clo.2}
  K(\kappa) = \mathtt{EllipticK}[\kappa^2],
  \quad
  E(\kappa) = \mathtt{EllipticE}[\kappa^2],  
\end{equation}
notice the exponents in the arguments. The result
(\ref{eq:examp.clo.1}) can be converted to spherical coordinates by
substituting $\rho=r\sin\theta, z=r\cos\theta$. Doing so leads to 
Jackson's textbook result as given in its eq.\ (5.37)~\cite{jack}.

On the other hand, the Laplace transform in (\ref{eq:examp.cyl.4}) is
evaluated in analytically closed form in eq.\ (13) of section 4.14 of
the table of integral transforms \cite{bates.lap}, which yields the
simpler expression, 
\begin{equation}
  \label{eq:examp.clo.3}
  A_\vphi(\bs{r}) = \frac{\mu_0 I}{2\pi}\sqrt{\frac{a}{\rho}}
  Q_{\frac{1}{2}}(\xi),
  \quad
  \xi=\frac{\rho^2+z^2+a^2}{2 a \rho}.
\end{equation}
In this equation $Q_\frac{1}{2}$ is a Legendre function of the second kind,
as defined in chapter III of \cite{bates.leg}. $Q_\frac{1}{2}(x)$ is
analytic in the complex $x$ plane cut along the real axis from
$-\infty$ to 1, and has a logarithmic branch point at $x=1$. Notice
that in (\ref{eq:examp.clo.3}) the argument is $\xi>1$ for
$\rho\neq a$ or $z\neq0$. It is only at $\rho=a$, $z=0$, where the
current loop is located, that $\xi=1$ and $Q_\frac{1}{2}(\xi)$ diverges. The
implementation of this function in Mathematica is given by
\begin{equation}
  \label{eq:examp.clo.4}
  Q_\nu(x) = Q_{\nu}^0(x) = \mathtt{LegendreQ[\nu,0,3,x]},
\end{equation}
where the argument ``3'' indicates the position of the cut.
As an illustration of the mathematical equivalence of eqs.\
(\ref{eq:examp.cyl.4}), (\ref{eq:examp.clo.1}),
(\ref{eq:examp.clo.3}), in table \ref{tab:tab} we give several numerical
values for $\frac{2\pi}{\mu_0 I} A_\vphi$ from those equations, with
the integral in (\ref{eq:examp.cyl.4}) and the special functions in (\ref{eq:examp.clo.1}),
(\ref{eq:examp.clo.3}) evaluated numerically with Mathematica. We find
perfect equality of the three values to six decimals at all values
of $\rho/a$, $z/a$, except for a minor numerical difference of order
$10^{-10}$ near the $z$ axis, where $A_\vphi$ vanishes.

\begin{table}
  \centering{}
  \begin{tabular}{cccccc}
   $\rho/a$ & $z/a$ & (\ref{eq:examp.cyl.4}) & (\ref{eq:examp.clo.1}) & (\ref{eq:examp.clo.3})\\\hline
    0.35 & 0.0 & 0.577159  &0.577159  & 0.577159 \\
    3.50 & 0.0 & 0.132367  &0.132367  & 0.132367 \\
    0.35 & 1.74& 0.065912  &0.065912  & 0.065912 \\    
    3.50 & 1.74& 0.091995  &0.091995  & 0.091995 \\
$10^{-6}$& 1.74& 1.94335$\times10^{-7}$&1.94335$\times10^{-7}$&1.94735$\times10^{-7}$\\\hline
  \end{tabular}
  \caption{Values of $\frac{2\pi}{\mu_0 I} A_\vphi$, for the
    indicated values of $\rho/a$, $z/a$, obtained from
    (\ref{eq:examp.cyl.4}), (\ref{eq:examp.clo.1}),
    (\ref{eq:examp.clo.3}) with \cite{wolf}.}
  \label{tab:tab}  
\end{table}

The equality between the two expressions (\ref{eq:examp.clo.1}) and
(\ref{eq:examp.clo.3}) for $A_\vphi(\bs{r})$ leads to a neat relation
between the Legendre function $Q_\frac{1}{2}$ and the complete
elliptic integrals $K$, $E$ that can be written as,
\begin{equation}
  \label{eq:examp.clo.6}
  Q_\frac{1}{2}(z) = \sqrt{\frac{2}{z+1}} \left(
    z K\left(\sqrt{\frac{2}{z+1}}\right) - (z+1) E\left(\sqrt{\frac{2}{z+1}}\right)
    \right),
\end{equation}
and can be easily checked numerically.

\subsection{Analytically closed expressions: magnetic induction field}
\label{sec:examp.analytic.mag.ind}

The magnetic induction field is obtained as
the curl of the vector potential (\ref{eq:examp.clo.3}) in cylindrical coordinates.
%\begin{equation*}
%\bs{B}(\bs{r}) = -\frac{\partial A_\varphi}{\partial z}(\bs{r})
%\bs{\hat{\rho}}
%+ \frac{1}{\rho} \frac{\partial(\rho A_\vphi)}{\partial\rho}(\bs{r}) \bs{\hat{z}}~.
%\end{equation*}
The partial derivatives can be expressed in terms of the derivative
\cite{bates.leg}
\begin{equation}
  \label{eq:examp.clo.5}
  \frac{d^m Q_\nu}{dz^m}(z) = (z^2-1)^{-\frac{m}{2}} Q_\nu^m(z),
  \quad m=1,2,3,\ldots,
\end{equation}
with $\nu=1/2$ and $m=1$. This way we are led to the magnetic induction field,
\begin{equation}
  \label{eq:examp.analitic.mag.ind.1}
  \begin{aligned}
  B_\rho &= -\frac{\partial A_\vphi}{\partial z}
        &\hspace*{-5pt}=& -\frac{\mu_0 I}{2\pi}\frac{2\sqrt{a/\rho}\,z}{\sqrt{(\rho-a)^2+z^2}\sqrt{(\rho+a)^2+z^2}}
           Q_\frac{1}{2}^1(\xi),\\
  B_z    &= \frac{1}{\rho}\frac{\partial}{\partial \rho} (\rho A_\vphi) 
        &\hspace*{-5pt}=& \frac{\mu_0 I}{2\pi} \frac{1}{\rho} \sqrt{\frac{a}{\rho}} \left(
           \frac{1}{2} Q_\frac{1}{2}(\xi) + \frac{\rho^2-a^2-z^2}{\sqrt{(\rho-a)^2+z^2}\sqrt{(\rho+a)^2+z^2}}
            Q_\frac{1}{2}^1(\xi) \right),
  \end{aligned}
\end{equation}
with $\xi$ as in (\ref{eq:examp.clo.3}). As a verification, by using
the power expansions as $\rho\rightarrow0$ \cite{bates.leg},
\begin{equation}
  \label{eq:examp.anal.mag.ind.2}
  Q_\frac{1}{2}(\xi) = \frac{\pi}{2} \frac{a^\frac{3}{2}}{(a^2+z^2)^\frac{3}{2}} \rho^\frac{3}{2} + O(\rho^\frac{7}{2}),
  \quad
  Q_\frac{1}{2}^1(\xi) = -\frac{3\pi}{4} \frac{a^\frac{3}{2}}{(a^2+z^2)^\frac{3}{2}} \rho^\frac{3}{2} + O(\rho^\frac{7}{2}),  
\end{equation}
which can also be obtained with Mathematica, we find the magnetic induction field on the $z$ axis,
\begin{equation}
  \label{eq:examp.anal.mag.ind.3}
  B_\rho(z\bs{\hat{z}})=0,
  \qquad
  B_z(z\bs{\hat{z}}) = \frac{\mu_0}{2\pi} \frac{I\pi a^2}{(a^2+z^2)^\frac{3}{2}}.
\end{equation}
This way we recover the result on the $z$ axis that is more usually obtained from the standard Amperian integral
\begin{equation*}
\bs{B}(\bs{r}) = \frac{\mu_0}{4\pi} \int d^3r' \bs{j}(\bs{r'}) \wedge \frac{\bs{r}-\bs{r'}}{|\bs{r}-\bs{r'}|^\frac{3}{2}}, 
\end{equation*}
with $\bs{r}=z\bs{\hat{z}}$ and $\bs{j}$ from (\ref{eq:examp.cyl.1}).

\section{Final remarks}
\label{sec:finrem}

In this brief note, we present eq.\ (\ref{eq:intro.6}) for the vector
potential of a steady-state azimuthal current density and provide an
elementary proof in section \ref{sec:vec.pot}. Our proof is simpler
than the spherical-coordinate derivation in \cite{vas12} and more
general than the cylindrical-coordinate approach in \cite{zang}. In
section \ref{sec:examp}, we examine the vector potential of a circular
current loop and recover its expansions in orthogonal eigenfunctions
in spherical coordinates \cite{jack} and in cylindrical ones
\cite{zang}. Furthermore, we derive a new, simpler
closed-form expression for this case, given by (\ref{eq:examp.clo.3})
in terms of the Legendre function of the second kind,
$Q_\frac{1}{2}$. This expression, together with
(\ref{eq:examp.clo.1}), leads to the relation (\ref{eq:examp.clo.6})
between this Legendre function and the complete elliptic integrals $E$
and $K$. By direct evaluation of the curl of (\ref{eq:examp.clo.3}) we
obtain the magnetic induction field $\bs{B}(\bs{r})$ of the circular
current loop in analytically closed form in
(\ref{eq:examp.analitic.mag.ind.1}) in cylindrical coordinates.

\end{document}